\documentclass[fleqn,twoside]{article}
\usepackage[headings]{espcrc2}

\newcommand{\be}{\begin{equation}}
\newcommand{\ee}{\end{equation}}
\newcommand{\ba}{\begin{eqnarray}}
\newcommand{\ea}{\end{eqnarray}}

\newcommand{\eq}{Eq.~}

\newcommand{\re}{Ref.~}
\newcommand{\nr}[1]{(\ref{#1})}
\newcommand{\nn}{\nonumber \\}
\newcommand{\fr}[2]{{\frac{#1}{#2}\,}}
\renewcommand{\(}{\left(}
\renewcommand{\)}{\right)}

\newcommand{\lk}{\left[}
\newcommand{\rk}{\right]}
\newcommand{\ld}{\left.}
\newcommand{\rd}{\right.}
\newcommand{\e}{\epsilon}
\newcommand{\ep}{\;\e}

\newcommand{\order}[1]{{\cal O}\(#1\)\vphantom{\fr12}}

\newcommand{\poch}[2]{{\(#1\)_{#2}}}
\newcommand{\pochi}[1]{{\(#1\)_{i}}}
\newcommand{\bFa}{{{}_2F_1}}
\newcommand{\cFb}{{{}_3F_2}}

\newcommand{\slnA}{\ln2}
\newcommand{\slnB}{\ln^22}
\newcommand{\slnC}{\ln^32}
\newcommand{\slnD}{\ln^42}
\newcommand{\slnE}{\ln^52}

\newcommand{\szB}{\zeta_2}
\newcommand{\szC}{\zeta_3}
\newcommand{\szE}{\zeta_5}
\newcommand{\szG}{\zeta_7}
\newcommand{\szI}{\zeta_9}

\newcommand{\saD}{a_4}
\newcommand{\saE}{a_5}

\newcommand{\sHa}{s_{8a}}

\newcommand{\sIa}{s_{9a}}

\newcommand{\bB}{{\rm B}_2}

\newcommand{\tC}{{\rm T}_3}

\def\-{\!-\!}
\def\+{\!+\!}
\def\ug{\!\!\!&=&\!\!\!}
\def\ue{\!\!\!&\equiv&\!\!\!}
\def\uu{\!\!\!&&\!\!\!}
\def\up{\!\!\!&+&\!\!\!}
\def\um{\!\!\!&-&\!\!\!}
\def\ut{\!\!\!&\times&\!\!\!}

\def\br{\rd\nn\uu{}\ld}

\title{%
Hypergeometric representation of a 
four-loop vacuum bubble}

\author{%
E.~Bejdakic\addressmark,
Y.~Schr\"oder\address{%
Fakult\"at f\"ur Physik, Universit\"at Bielefeld, 
33501 Bielefeld, Germany}
}

\begin{document}

\begin{abstract}
In this article, we present a new analytic result for a certain
single-mass-scale 
four-loop vacuum (bubble) integral. 
We also discuss its systematic $\e$\/-expansion
in $d=4-2\e$ as well as $d=3-2\e$ dimensions, 
the coefficients of which are expressible
in terms of harmonic sums at infinity.
\vspace{1pc}
\end{abstract}

\maketitle


\section{\label{se:intro}Introduction}

Recently, a number of calculations have reached the four-loop
level \cite{4loopExamples}. 
In most cases, such high-loop computations proceed by first 
reducing the (typically very large number of) momentum-space 
integrals to a small number of so-called master integrals,
a step that can be highly automatized \cite{Reduction}. 
In a second step, these master integrals have to be computed
in an $\e$\/-expansion around the space-time dimension $d$ of
interest.

This second step is much less amenable to automatization, at least
if one is interested in analytic expressions for the coefficients.
Indeed, very few four-loop master integrals are known analytically. 
It is the purpose of this paper to add one master integral 
to this set.

As a basic building block, many of the abovementioned calculations
use single-mass-scale vacuum bubbles. 
These arise e.g. when there are large scale hierarchies in the physics
problem, such that asymptotic expansions can be used.
Having no external momenta, and propagators with masses $0$ or $m$ only,
members of this class of integrals are functions of the space-time dimension
$d$ only (for convenience, we set $m=1$ in the remainder of this paper),
hence the coefficients of their Laurent expansions in $\e$ are pure numbers.

Define the (dimensionless, but $d$\/-dependent) 
1-, 3- and 4-loop integrals 
\ba 
J(x) \ue \int_{p}^{(d)}\frac1{\(p^2\+1\)^x} \;,\nn
\bB(x) \ue \int_{p_{1,2,3}}^{(d)}
\frac1{\(p_1^2\+1\)^x}\,\frac1{p_2^2\+1}\,
\frac1{p_3^2}\,\frac1{(p_1\+p_2\+p_3)^2} \nn
\tC(x) \ue
\int_{p_{1,2,3,4}}^{(d)}
\frac1{\(p_1^2\+1\)^x}\,\frac1{(p_1\+p_4)^2}\,
\frac1{p_2^2\+1}
\times\nn&&\hphantom{xxx}\times
\frac1{(p_2\+p_4)^2}\,
\frac1{p_3^2\+1}\,\frac1{(p_3\+p_4)^2} \;.
\nonumber
\ea
At $x=1$, all three examples are master integrals.
While $J(1)$ and $\bB(1)$ can be computed analytically in terms
of Gamma functions, we would like to know $\tC(1)$.


\section{\label{se:deq}Difference equation}

Using a slight generalization of the abovementioned 
reduction step, one can systematically 
derive sets of difference equations for the integrals.
For an integral $f(x)$,
they have the generic form $\sum_{i=0}^N c_i f(x+i)=g(x,d)$,
with polynomial coefficients $c_i(d)$, and
where $N$ is the {\em order} of the difference equation
and $g(x)$ its inhomogeneity,
which is supposed to be known analytically.

First-order difference equations can be solved directly
in terms of one initial value. 

In the homogeneous case, this can be done trivially in terms
of Pochhammer symbols $\poch{a}{x}\equiv\Gamma\(a+x\)/\Gamma\(a\)$:
\ba
J(x+1) 
\ug  \frac{x-\frac{d}2}x\, J(x) \nn
\ug  \frac{\poch{1-\frac{d}2}{x}}{\poch{1}{x}}\, J(1) \;, \\
\bB(x+1) 
\ug  \frac{\(x+2-d\)\(x+3-\frac{3d}2\)}{x\(x+5-2d\)}\, \bB(x) \nn
\ug  \frac{\poch{3-d}{x} \poch{4-\frac{3d}2}{x}}{\poch{1}{x} 
\poch{6-2d}{x}}\, \bB(1) \;.
\ea
For these two cases, we know the initial values $J(1)$ and $\bB(1)$ 
analytically in $d$ dimensions.
The value of $J(1)$ depends on the choice of integration
measure, which we do not need to specify here\footnote{To display one
particular choice, $J(1)=\int{\rm d}^dk/(k^2+1)=\pi^{d/2}\Gamma(1-d/2)$.}, 
and a simple integration gives
\ba
\bB(1) \ug 
\frac{2^{d\-3}\Gamma\(4\-\frac{3d}2\)\Gamma\(\frac32\-\frac{d}2\)
\Gamma\(\frac{d}2\)}
{\Gamma\(\frac72-d\)\Gamma\(1-\frac{d}2\)}\lk J(1)\rk^3 \;.
\ea

The 4-loop integral $\tC$ is an example for the inhomogeneous case,
\ba
\tC(x+1) \ug  \frac{(x+2-d)(x+5-2d)}{x(x+8-3d)}\, \tC(x) +
\nn\up {}\frac1{x\+8\-3d} \lk 
(d\-2)J(1)\bB(x+1) \vphantom{\frac12} +
\rd\nn\up {}\ld 
 \!\!\(4-\frac{3d}2\) \bB(1) J(x+1)\rk \\
\label{eq:t3deq}
\ue c(x) \tC(x)+G(x) \;.
\ea
This difference equation has been solved numerically 
in 3d \cite{Schroder:2003kb} and 4d \cite{Schroder:2005va}.
We will solve this equation analytically in the following section.


\section{\label{se:soln}Solution}

One can immediately write down a solution of \eq\nr{eq:t3deq} in terms of a 
sum, as can be easily seen by doing the first couple of iterations,
\ba
\label{eq:t3soln}
\tC(x+1) \ug  \tC(x_0) \( \prod_{i=x_0}^x c(i) \)
+\nn\uu {} 
+\sum_{j=x_0}^x G(j) \( \prod_{i=j+1}^x c(i) \) \;.
\ea
In this case however, it is the initial value $\tC(1)$ that 
we would like to compute. What we instead know about the integral $\tC(x)$
is its behavior at the boundary, $\tC(x\gg 1) \propto J(x\gg 1)$.

To proceed, let us rewrite \eq\nr{eq:t3soln} as
\ba
\tC(x_0) \ug  \tC(x+1) \( \prod_{i=x_0}^x \frac1{c(i)} \)
-\nn\uu {}
-\sum_{j=x_0}^x G(j) \( \prod_{i=x_0}^j \frac1{c(i)} \) \\
\ug  \lk \tC(x\+1) \frac{\Gamma(x+1)\Gamma(x+9-3d)}
{\Gamma(x\+3\-d)\Gamma(x\+6\-2d)}
- \vphantom{\sum_{j=x_0}^x}\rd\nn\uu{} \ld
\-\sum_{j=x_0}^x\!\! G(j) \frac{\Gamma(j+1)\Gamma(j+9-3d)}
{\Gamma(j\+3\-d)\Gamma(j\+6\-2d)} \rk 
\times\nn\uu {}\times
\frac{\Gamma(x_0+2-d)\Gamma(x_0+5-2d)}{\Gamma(x_0)\Gamma(x_0+8-3d)} \;.
\label{eq:T3.x0}
\ea
The left-hand side does not depend on $x$, so the same has to be true 
for the right-hand side.
Hence, we can freely choose the value of $x$ at which we want to evaluate
the right-hand side; in particular we can choose $x\gg 1$ or even perform  
the limit
$\lim_{x\rightarrow\infty}$.

Using Stirling's formula 
$\Gamma(x+a)/\Gamma(x+b) \stackrel{x\gg 1}{=}x^{a-b}\(1+{\cal O}\(x^{-1}\)\)$,
we see that the first term on the right-hand side of \eq\nr{eq:T3.x0}
behaves like $x^{1-d/2}$ at 
large $x$, so it vanishes for $d>2$. So we are left to evaluate
\ba
\tC(x_0) \!\!\!&\stackrel{d>2}{=}&\!\!\! 
-\frac{\Gamma(x_0+2-d)\Gamma(x_0+5-2d)}{\Gamma(x_0)\Gamma(x_0+8-3d)}
\times\nn\ut\!\!\!
\sum_{j=0}^{\infty} \!\frac{\Gamma(j\+x_0\+1)\Gamma(j\+x_0\+8\-3d)}
{\Gamma(j\+x_0\+3\-d)\Gamma(j\+x_0\+6\-2d)} 
\!\times\nn\ut\!\!\!
\lk (d-2) \frac{\bB(j+x_0+1)}{\bB(1)} 
+\rd\nn\up\ld\!\!\!
\frac{8-3d}2\, 
\frac{J(j\+x_0\+1)}{J(1)} \rk J(1) \bB(1) \;.
\ea
Multiplying by $\poch{1}{j}/\poch{1}{j}$, the sum is just 
a generalized hypergeometric function ${}_{J+1}F_{J}$ at
unit argument,
\ba
\tC(x_0) \!\! \!\!\!&\stackrel{d>2}{=}&\!\!\! \! 
\frac{\Gamma(x_0+2-d)\(4-\frac{3d}2\)}
{(x_0\+5\-2d)\Gamma(x_0)\Gamma(2\-d)}\, J(1) \bB(1) 
\!\times\nn\ut\!\!\!
\lk 
\cFb\!\(\!i_1,i_2,1;i_3,i_3;1\!\) 
\frac{\Gamma(i_1)\Gamma(6-2d)}
{\Gamma\(5\-\frac{3d}2\)\Gamma(i_3)} -
\rd\nn\um\!\!\!\ld
\cFb\!\(i_4,i_2,1;i_5,i_3;1\) 
\frac{\Gamma(i_4)\Gamma(2-d)}
{\Gamma\(1\-\frac{d}2\)\Gamma(i_5)}
\rk
\nonumber
\ea
where 
$i_1=x_0+4-\frac{3d}2$,
$i_2=x_0+8-3d$,
$i_3=x_0+6-2d$,
$i_4=x_0+1-\frac{d}2$ and
$i_5=x_0+3-d$.

From this, setting $x_0=1$, we finally get \cite{ervin}
\ba\label{eq:finally}
\tC(1) \!\!\!&\stackrel{d>2}{=}&\!\!\!  
\frac{(2-d)(8-3d)}{8(3-d)^2}\, J(1) \bB(1) 
\times\nn\ut\!\!\! \!\!
\lk 
\cFb\!\(\!5\-\frac{3d}2,9\-3d,1;7\-2d,7\-2d;1\!\) \-
\rd\nn\um\!\!\! \!\!\ld
\cFb\!\(\!2\-\frac{d}2,9\-3d,1;4\-d,7\-2d;1\!\) 
\rk \;.
\ea
We will now turn to expanding this function around $d=4$ and $d=3$.


\section{\label{se:4d}Expansion in 4d}

Following the notation of \re\cite{Schroder:2005va}, we define
\ba
\tC \ue \frac{\tC(1)}{[J(1)]^4} \;.
\ea
For $d=4-2\e$, \eq\nr{eq:finally} then reads
\ba
\tC \ug  \frac1{2^{2\e}}\, \frac{1\-\e}{(1\-2\e)^2}\,
\frac{\Gamma\(3\e\-1\)\Gamma\(\e\-\frac12\)\Gamma\(2\-\e\)}
{\Gamma\(2\e\-\frac12\)\Gamma\(\e\-1\)}
\times\nn\ut\!\! 
\lk \cFb\(1,6\e\-3,\e;4\e\-1,2\e;1\) 
-\rd\nn\um\!\!\ld
\cFb\(1,6\e\-3,3\e\-1;4\e\-1,4\e\-1;1\) \rk
\ea

An $\ep$\/-expansion of the $\cFb$
can be achieved via Algorithm A of \re\cite{Moch:2001zr}.
We have used the package XSummer \cite{Moch:2005uc} 
to expand in $\e$, in terms of harmonic polylogarithms 
$H(z)$ \cite{Remiddi:1999ew} of unit argument.
In a second step, we have used the package Summer \cite{table9} 
to rewrite the $H(1)$ in terms of a minimal set of numbers,
which are equivalent to harmonic sums $S_{\vec m}(n)$ 
at infinity \cite{Vermaseren:1998uu}.

We have coded both expansions in FORM \cite{form},
and obtain (note the absence of an $\e^2$ term)
\ba \label{eq:T3.4d.inc}
{   }
{    \tC } \ug {  }
{   }
{ + \frac{ 1 }{ 4 } }
{   }
{        + \e } \(
{ + \frac{ 1 }{ 2 } }
   \) 
{   }
{        + \e^3 } \(
{           - 8  }
{ + \frac{ 13 }{ 2 }{ \,\szC } }
   \) \nn\uu{}
{   }
{        + \e^4 } \(
{ - \frac{ 241 }{ 4 } }
{ - \frac{ 45 }{ 2 }{ \,\szB^2 } }
{           + 4 }{ \,\szC  }
   \) \nn\uu{}
{   }
{        + \e^5 } \(
{ - \frac{ 669 }{ 2 } }
{ - \frac{ 36 }{ 5 }{ \,\szB^2 } }
{           - 36 }{ \,\szC  }
{ + \frac{ 693 }{ 2 }{ \,\szE } }
   \) \nn\uu{}
{   }
{        + \e^6 } \(
{           - 1636  }
{ + \frac{ 756 }{ 5 }{ \,\szB^2 } }
{ - \frac{ 3168 }{ 7 }{ \,\szB^3 } }
{           - 289 }{ \,\szC  }
\br
{ + \frac{ 241 }{ 2 }{ \,\szC^2 } }
{           + 72 }{ \,\szE  }
   \) \nn\uu{}
{   }
{        + \e^7 } \(
{           - 7472  }
{ + \frac{ 5301 }{ 5 }{ \,\szB^2 } }
{ - \frac{ 432 }{ 7 }{ \,\szB^3 } }
{ - \frac{ 3061 }{ 2 }{ \,\szC } }
\br
{ - \frac{ 4437 }{ 5 }{ \,\szC }{ \,\szB^2 } }
{           + 16 }{ \,\szC^2  }
{           - 2484 }{ \,\szE  }
{ + \frac{ 45921 }{ 4 }{ \,\szG } }
   \) \nn\uu{}
{   }
{        + \e^8 } \(
{           - 32736  }
{ + \frac{ 52713 }{ 10 }{ \,\szB^2 } }
{ + \frac{ 23616 }{ 7 }{ \,\szB^3 } }
\br
{ - \frac{ 13749777 }{ 1750 }{ \,\szB^4 } }
{           - 7059 }{ \,\szC  }
{ - \frac{ 288 }{ 5 }{ \,\szC }{ \,\szB^2 } }
{           - 900 }{ \,\szC^2  }
\br
{           - 16677 }{ \,\szE  }
{           + 13977 }{ \,\szE }{ \,\szC  }
{           + 996 }{ \,\szG  }
{ + \frac{ 5238 }{ 5 }{ \,\sHa } }
   \) \nn\uu{}
{   }
{        + \e^9 } \(
{           - 139604  }
{ + \frac{ 116127 }{ 5 }{ \,\szB^2 } }
{ + \frac{ 154512 }{ 7 }{ \,\szB^3 } }
\br
{ - \frac{ 11232 }{ 25 }{ \,\szB^4 } }
{           - 30584 }{ \,\szC  }
{ + \frac{ 34344 }{ 5 }{ \,\szC }{ \,\szB^2 } }
\br
{ - \frac{ 129600 }{ 7 }{ \,\szC }{ \,\szB^3 } }
{           - 5881 }{ \,\szC^2  }
{ + \frac{ 4933 }{ 3 }{ \,\szC^3 } }
\br
{ - \frac{ 162045 }{ 2 }{ \,\szE } }
{           - 30051 }{ \,\szE }{ \,\szB^2  }
{           + 576 }{ \,\szE }{ \,\szC  }
\br
{           - 87858 }{ \,\szG  }
{ + \frac{ 2094913 }{ 6 }{ \,\szI } }
   \) 
{   }
{        + } \order { \e^{10} }  
\;,
\ea
where $\zeta_n=\zeta(n)$ are values of the Riemann Zeta function, 
and $\sHa= S_{5,3}(\infty)\approx1.041785$ is the only non-zeta 
value appearing up to this order.
Up to the $\e^5$ term, these coefficients have already been
obtained (by fitting high-precision numerical values) 
in \re\cite{Schroder:2005va}. Higher-order coefficients are new.
To 30 digit accuracy, \eq\nr{eq:T3.4d.inc} reads
\ba
\tC \ug{} + 
  0.250000000000000000000000000000 
\nn \uu{} + 
  0.500000000000000000000000000000 \,\e
\nn \uu{} - 
  0.186630129462637144901701950176 \,\e^{3}
\nn \uu{} - 
  116.322454283613146131176255284 \,\e^{4}
\nn \uu{} - 
  37.9603995633682022478604992948 \,\e^{5}
\nn \uu{} - 
  3339.84136081794624883976306535 \,\e^{6}
\nn \uu{} - 
  580.540498497887663112519574990 \,\e^{7}
\nn \uu{} - 
  68729.9111560055991193326976118 \,\e^{8}
\nn \uu{} - 
  7274.24628251698563686391222230 \,\e^{9}
\nn \uu{} 
+ \order{\e^{10}} \;. \nonumber
\ea


\section{\label{se:3d}Expansion in 3d}

For $d=3-2\e$, \eq\nr{eq:finally} reads
\ba
\tC \ug  \frac1{4^{1\+\e}}\, \frac{1\-2\e}{4\e^2}
\frac{\Gamma\(\frac12\+3\e\)\Gamma\(\e\)\Gamma\(\frac32\-\e\)}
{\Gamma\(\frac12\+2\e\)\Gamma\(-\frac12\+\e\)}
\,\times\nn\ut\!\!{} 
\lk \cFb\(1,6\e,\frac12+\e;1+4\e,1+2\e;1\) 
-\rd\nn\um\!\!\ld
\cFb\(1,6\e,\frac12+3\e;1+4\e,1+4\e;1\) \rk
\ea
XSummer cannot expand the $\cFb$ around half-integer indices, so 
unfortunately we cannot proceed the same way as above. 
There are algorithms for expansion around rationals p/q
\cite{Weinzierl:2004bn},
which work only if rationals are {\em balanced} between numerator 
and denominator.
The case at hand is {\em unbalanced}.

To balance it, we may make use of the Euler identity for Gau\ss{}' 
hypergeometric function
\ba
\bFa(\alpha,\beta;\delta;x) \ug  (1\-x)^{\delta\-\alpha\-\beta}
\bFa(\delta\-\alpha,\delta\-\beta;\delta;x) 
\nonumber
\ea
which allows us to rewrite
\ba
&&\hspace*{-10mm} \cFb(\alpha,\beta,\gamma;\delta,\e;1) 
\;=\nn\ug 
\frac{\Gamma(\e)}{\Gamma(\gamma)\Gamma(\e\-\gamma)}
\!\int_0^1\!\!\!{\rm d}x\, \frac{(1\-x)^{\e\-\gamma\-1}}{x^{1-\gamma}}\,
\bFa(\alpha,\beta;\delta;x) \nn
\ug  \frac{\Gamma(\e)\Gamma(\e+\delta-\alpha-\beta-\gamma)}
{\Gamma(\e-\gamma)\Gamma(\e+\delta-\alpha-\beta)}
\,\times\nn\ut
\cFb(\delta-\alpha,\delta-\beta,\gamma;\delta,\e+\delta-\alpha-\beta;1) \;.
\ea

This gives 6 transforms, 4 of which are {\em balanced}. 
Choosing one of those possibilities\footnote{Note that both 
generalized hypergeometric functions we need can be conveniently 
summarized introducing only one parameter, $a=2$ and $a=4$, respectively.}, 
\ba
&&\hspace*{-10mm} \cFb\(1,\frac12+(a-1)\e,6\e;1+a\e,1+4\e;1\) 
\;=\nn\ug 
\frac{\Gamma(1+4\e)\Gamma(\frac12-\e)}{\Gamma(1-2\e)\Gamma(\frac12+5\e)}
\,\times\nn\ut
\cFb\(a\e,\frac12+\e,6\e;1+a\e,\frac12+5\e;1\) 
\nn \ug 
\frac{\Gamma(1+4\e)\Gamma(\frac12-\e)}{\Gamma(1-2\e)\Gamma(\frac12+5\e)}
\,\times\nn\ut
\(1+\sum_{i=1}^\infty \frac{\pochi{a\e}\pochi{\frac12+\e}\pochi{6\e}}
{\pochi{1+a\e}\pochi{\frac12+5\e}\pochi{1}} \) \;,
\ea
where $\pochi{n}=\frac{\Gamma(i+n)}{\Gamma(n)}$ are Pochhammer symbols.
The sum is of ``type A'' according to the classification in 
\cite{Weinzierl:2004bn},
a class that can be expanded in terms of harmonic sums, as we will show now.

Using $\pochi{\e}=-\sum_{n=1}^\infty \(-\frac{\e}{i}\)^n\,\pochi{1+\e}$ 
for two terms in the numerator, $\pochi{a\e}$ and $\pochi{6\e}$,
and solving one of the two new sums, the above sum reads
\ba
\sum_{n=2}^\infty \frac{(-6\e)^n a}{6\-a}\(1\-\(\frac{a}6\)^{n\-1}\)
\sum_{i=1}^\infty \frac{\pochi{\frac12\+\e}\pochi{1\+6\e}}
{i^n\pochi{\frac12\+5\e}\pochi{1}} \;. \nonumber
\ea

The next step is to expand the Pochhammer symbols in $\e$, using 
\cite{Weinzierl:2004bn}
\ba
&&\hspace*{-10mm} \pochi{\frac12+\e} 
\;=\nn\ug\!\!
\pochi{\frac12} \exp\(\!{-\sum_{k=1}^\infty 
\frac{(-2\e)^k}{2k}\lk S_k(2i)\-S_{-k}(2i) \rk}\!\) \nn
&&\hspace*{-10mm} \pochi{1+\e} 
\;=\nn\ug\!\! 
\pochi{1} \exp\({-\sum_{k=1}^\infty 
\frac{(-2\e)^k}{2k}\lk S_k(2i)\+S_{-k}(2i) \rk}\) \;.\nonumber
\ea
It becomes clear now why the sum had to be {\em balanced}: the Pochhammer
symbols $\pochi{\frac12}$ and $\pochi{1}$ cancel between numerator 
and denominator. 
Expanding the exponentials and rewriting products of S-sums as
single S-sums leaves us with sums of generic form 
$\sum_i S_{\vec k}(2i)/i^n$ to be solved.

To this end, we use the trick employed in \cite{Weinzierl:2004bn},
introducing a delta function on the integers:
\ba
\sum_{i=1}^\infty \frac{S_{\vec k}(2i)}{i^n} 
\ug 
2^n \sum_{i=1}^\infty \frac1{(2i)^n} S_{\vec k}(2i)
\nn\ug
\frac{2^n}2 \sum_{j=1}^\infty \frac{1^j+(-1)^j}{j^n} S_{\vec k}(j)
\nn\ug
\frac{2^n}2 \lk S_{n,\vec k}(\infty) +S_{-n,\vec k}(\infty) \rk \;.
\ea

Putting it all together, and doing some trivial algebra with Gamma functions,
\ba \label{eq:T3.3d.exp}
\tC \ug  3 \frac{2^{6\e}}{16\e}  (1\-2\e)^3 
\frac{\Gamma^2(1\+\e)}{\Gamma(1\-\e)}\,
\frac{\Gamma(\frac12\+3\e)\Gamma(\frac12\-\e)}
{\Gamma(\frac12+5\e)\Gamma(\frac12)}
\!\times\nn\ut\!\!\!
\sum_{n=0}^\infty (-4\e)^n \(1\+3^{n\+2}\-2^{n\+3}\) 
\sum_{j=1}^\infty \frac{1^j\+(-1)^j}{j^{n+2}} 
\!\times\nn\ut\!\!
\exp\(\sum_{k=1}^\infty \frac{(-2\e)^k}{(-2k)} 
\lk (6^k-5^k+1)S_k(j)
+\rd\rd\nn\uu\!\!\ld\ld
+(6^k+5^k-1)S_{-k}(j)\rk \vphantom{\sum_{k=1}^\infty} \) \;.
\ea

For expanding the Gamma functions around integer and half-integer 
arguments, we use
\ba
\Gamma\(1+\e\) \ug  
\exp\(-\e\gamma+\sum_{n\ge2} (-\e)^n\frac{\zeta(n)}{n}\),\\
\Gamma\(\frac12+\e\) \ug  \sqrt{\pi}\,
\exp\(-\e\(\gamma+2\ln2\)
+\vphantom{\sum_{n\ge2}(-\e)^n\frac{\zeta(n)}{n}}\rd\nn\uu\ld
+\sum_{n\ge2}(-\e)^n\frac{\zeta(n)}{n}\(2^n-1\)\) \;,
\ea
such that \eq\nr{eq:T3.3d.exp} gives
\ba
{   }
{    \tC } \ug {  }
{   }
{        + \e }^{ -1 } \(
{ + \frac{ 3 }{ 16 }{ \,\szB } }
   \) 
\nn\uu{}
{   }
{   + } \(
{ - \frac{ 9 }{ 8 }{ \,\szB } }
{ + \frac{ 9 }{ 4 }{ \,\szB }{ \,\slnA } }
{ - \frac{ 21 }{ 8 }{ \,\szC } }
  \) \nn\uu{}
{   }
{        + \e } \(
{ + \frac{ 3 }{ 2 }{ \,\slnD } }
{ + \frac{ 9 }{ 4 }{ \,\szB } }
{ - \frac{ 27 }{ 2 }{ \,\szB }{ \,\slnA } }
{ + \frac{ 9 }{ 2 }{ \,\szB }{ \,\slnB } }
\br
{ - \frac{ 207 }{ 40 }{ \,\szB^2 } }
{ + \frac{ 63 }{ 4 }{ \,\szC } }
{           + 36 }{ \,\saD  }
   \) \nn\uu{}
{   }
{        + \e^2 } \(
{           - 9 }{ \,\slnD  }
{ + \frac{ 18 }{ 5 }{ \,\slnE } }
{ - \frac{ 3 }{ 2 }{ \,\szB } }
{           + 27 }{ \,\szB }{ \,\slnA  }
\br
{           - 27 }{ \,\szB }{ \,\slnB  }
{           + 18 }{ \,\szB }{ \,\slnC  }
{ + \frac{ 621 }{ 20 }{ \,\szB^2 } }
\br
{ - \frac{ 621 }{ 10 }{ \,\szB^2 }{ \,\slnA } }
{ - \frac{ 63 }{ 2 }{ \,\szC } }
{ - \frac{ 87 }{ 8 }{ \,\szC }{ \,\szB } }
{ + \frac{ 4743 }{ 16 }{ \,\szE } }
\br
\vphantom{\frac12}
{           - 216 }{ \,\saD  }
{           - 432 }{ \,\saE  }
   \) 
{   }
{   }
{        + } \order { \e^3 }  
\;,
\ea
where $a_n={\rm Li}_n\(1/2\)$ are polylogarithms.
Due to space limitations, we have not shown all known coefficients
above. For the full result including $\order{\e^6}$ 
(whose coefficient entails weight-9 numbers like $\sIa$), see \cite{YSprep}.
To 30 digit accuracy, we have
\ba
\tC \!\ug{} + 
  0.308425137534042456838577843746 \,\e^{-1}
\nn \uu{} - 
  2.44054202205177103588797505371 
\nn \uu{} + 
  15.7704017490035557236803526639 \,\e
\nn \uu{} - 
  100.504895668775469387159215590 \,\e^{2}
\nn \uu{} + 
  625.061590652357317030007023983 \,\e^{3}
\nn \uu{} - 
  3839.50600368202197971416339398 \,\e^{4}
\nn \uu{} + 
  23392.0898313583063340598037114 \,\e^{5}
\nn \uu{} - 
  141772.262976662567133066547299 \,\e^{6}
\nn \uu{} + \order{\e^{7}} \;. \nonumber
\ea


\section*{Acknowledgments}

We would like to thank S.~Moch and S.~Weinzierl for communication.


\end{document}